\documentclass[conference]{IEEEtran}
\IEEEoverridecommandlockouts
\usepackage{amsmath}
\DeclareMathOperator{\Tr}{Tr}
\usepackage{cite}
\usepackage{amsmath,amssymb,amsfonts}
\usepackage{algorithmic}
\usepackage{algorithm}
\usepackage{graphicx}
\usepackage{textcomp}
\usepackage{xcolor}
\usepackage{stfloats}
\usepackage{url}
\usepackage{subfig}
\usepackage{graphicx}
\usepackage{bm}
\usepackage{booktabs}
\usepackage{cite}
\usepackage{color}
\usepackage{makecell}
\def\BibTeX{{\rm B\kern-.05em{\sc i\kern-.025em b}\kern-.08em
    T\kern-.1667em\lower.7ex\hbox{E}\kern-.125emX}}
\begin{document}

\title{Synesthesia of Machines (SoM)-Aided FDD Precoding with Sensing Heterogeneity: A Vertical Federated Learning Approach\\


}

\author{\IEEEauthorblockN{Haotian Zhang\IEEEauthorrefmark{1}, Shijian Gao\IEEEauthorrefmark{2}\IEEEauthorrefmark{3}, Weibo Wen\IEEEauthorrefmark{1} and Xiang Cheng\IEEEauthorrefmark{1} }

\IEEEauthorblockA{\IEEEauthorrefmark{1}The State Key Laboratory of Photonics and Communications, School of Electronics, \\Peking University, Beijing,  P. R. China} \IEEEauthorblockA{\IEEEauthorrefmark{2}The Internet of Things Thrust, The Hong Kong University of Science and Technology (Guangzhou),\\ Guangzhou, P. R. China} 
\IEEEauthorblockA{\IEEEauthorrefmark{3}Guangdong Provincial Key Laboratory of Future Networks of Intelligence, The Chinese University of Hong Kong,\\ Shenzhen, P. R. China} 
Emails: \{haotianzhang, weber\}@stu.pku.edu.cn, shijiangao@hkust-gz.edu.cn, xiangcheng@pku.edu.cn
	\thanks{

	This work was supported in part by the National Natural Science Foundation of China (Grants No. 62125101, 62341101, and 62401488), in part by the the New Cornerstone Science Foundation through the XPLORER PRIZE,
	and in part by the Guangdong Provincial Key Laboratory of Future Networks of Intelligence, The Chinese University of Hong Kong, Shenzhen (No. 2022B1212010001-OF08). (\textit{Corresponding Authors: Xiang Cheng; Shijian Gao.})
	
}
\\
}

\maketitle

\begin{abstract}
High complexity in precoding design for frequency division duplex systems necessitates streamlined solutions. Guided by Synesthesia of Machines (SoM), this paper introduces a heterogeneous multi-vehicle, multi-modal sensing aided precoding scheme within a vertical federated learning (VFL) framework, which significantly minimizes pilot sequence length while optimizing the system's sum rate. We address the challenges posed by local data heterogeneity due to varying on-board sensor configurations through a meticulously designed VFL training procedure. To extract valuable channel features from multi-modal sensing, we employ three distinct data preprocessing methods that convert raw data into informative representations relevant for precoding. Additionally, we propose an online training strategy based on VFL framework, enabling the scheme to adapt dynamically to fluctuations in user numbers. Numerical results indicate that our approach, utilizing short pilot sequences, closely approximates the performance of traditional optimization methods with perfect channel state information.
\end{abstract}

\section{Introduction}
Massive multiple-input multiple-output (mMIMO) is a key performance booster for 5G and future generation wireless networks \cite{mMIMO1,mMIMO2}. To mitigate inter-user interference and enhance throughput, it is essential to carefully consider the precoding design at base station (BS). In frequency-division duplexing (FDD) systems, BS precoding necessitates downlink channel estimation and uplink feedback to acquire downlink channel state information (CSI). This process incurs long latency under mMIMO, eroding the potential benefits of FDD systems. Furthermore, centralized procoding methods relying on downlink CSI are typically time-consuming due to high complexity caused by high dimensionality of CSI and complex matrix computation. Hence, optimizing precoding while reducing pilot overhead and computational complexity becomes a challenging problem, holding significant importance for latency-sensitive applications like vehicular networks \cite{VCN,FanV2V}.

Extensive endeavors have been dedicated to precoding optimization. The commonly adopted linear precoding approaches such as the zero-forcing (ZF) \cite{ZF} and the weighted minimum mean square error (WMMSE) method \cite{WMMSE} can offer outstanding performance under the premise of known perfect CSI. Nonetheless, the associated high pilot overhead and the exceedingly high complexity are intolerable in practical. Such issues are also faced by many centralized optimization schemes \cite{opt-1,opt-2}. 
Towards pilot overhead reduction, compressed sensing based methods leveraging channel sparsity have been extensively studied \cite{CS1,CS2}. However, they primarily rely on sparsity as prior information, and their {\color{black}computational} complexity is high since solving compressed sensing problems typically involves iterative algorithms. To further optimize the precoding performance in specific scenarios, recent works have proposed to utilize deep learning for FDD precoding \cite{Weiyu-MISO,Weiyu-MIMO}. Considering the significant time, workforce, and spectral resource costs and privacy issue associated with dataset collection in centralized learning (CL), researchers proposed to replace CL with the federated learning (FL) framework to optimize beamforming performance in a distributed manner \cite{beamformingFL3}. Yet, we observe that both CL and FL schemes still require a long downlink pilot sequence to approach the performance of ZF or WMMSE schemes. A new perspective from the non-radio frequency domain to decrease pilot overhead is offered by \cite{b3, MMFF}, where multi-modal sensory data collected by vehicles and BS is both proved to be effective in partially realizing the role of pilots by perceiving communication environment. However, the beam selection task addressed by \cite{b3,MMFF} has not been extended to BS precoding yet and the heterogeneity of on-board sensing is not taken into account.

In this work, we propose a heterogeneous multi-vehicle, multi-modal (H-MVMM) sensing aided precoding method, aiming to enhance system sum rate with reduced complexity and pilot overhead. We consider a crucial fact that vehicles in practical applications are equipped with different sensors, allowing them to capture unique communication environment features from different viewpoints.  To fully leverage the diversity of the multi-view features, we adopt the vertical FL (VFL) framework since VFL is tailored for situations where users have distinct feature spaces but share the same sample space (i.e, BS precoding).  A benefit offered by VFL is the reduction in computational complexity. This is achieved by dispersing the centralized precoding optimization from BS to distributed users, who only need to compute their own precoding vectors based on low-dimensional inputs. To bridge the substantial gap between multi-modal sensing and precoding, we customize three unique data preprocessing methods to transform raw sensory data into precoding-related informative representations guided by Synesthesia of Machines (SoM) \cite{SoM}. To mitigate the effects of data heterogeneity on VFL training, we design a customized loss function that can accelerate local model convergence. Additionally, we devise an online training strategy based on VFL framework, enabling the proposed scheme to dynamically adjust to user number changes. 

\section{System Model and Problem Formulation}
We consider an FDD multi-user MIMO system where a BS is equipped with an $ N_{\rm v} \times N_{\rm h}$ uniform planar array serving $K$ single-antenna users. $N_{\rm v}$ and $N_{\rm h}$ denote the numbers of vertical and horizontal antennas, with $N=N_{\rm v} N_{\rm h}$. We assume that the BS employs linear precoding to transmit the data symbols $s_k$ for each user. This can be expressed as follows:
\begin{equation}
\label{V}
\mathbf{x} = \sum^K_{k=1}\mathbf{v}_k s_k = \mathbf{V}\mathbf{s},
\end{equation}
where {\color{black}$\mathbf{v}_k \in \mathbb{C}^N$} represents the precoding vector for the $k$-th user and is also the $k$-th coloumn of the precoding matrix $\mathbf{V} \in \mathbb{C}^{N \times K}$. The precoding matrix $\mathbf{V}$ satisfies the transmit power constraint $P$ as $\Tr(\mathbf{V}\mathbf{V}^H) \leq P $. The power of data symbols is also normalized so that $\mathbb{E}[\mathbf{s}\mathbf{s}^H]=\mathbf{I}$. 

Let $\mathbf{h}_k \in \mathbb{C}^{N}$ denote the channel between BS and the $k$-th user. The received signal at the $k$-th user in downlink data transmission stage can be expressed as $y_k = \mathbf{h}_k^H\mathbf{v}_k s_k + \sum_{i \neq k}\mathbf{h}_k^H\mathbf{v}_i s_i + n_k$, where $n_k \sim \mathcal{CN}(0,\sigma^2)$ denotes additive white Gaussian noise. According to the received signal model, the achievable rate of the $k$-th user can be expressed as:
\begin{equation}
\label{Y}
R_k = \log_2(1+\frac{|\mathbf{h}_k^H\mathbf{v}_k|^2}{\sum_{i \neq k}|\mathbf{h}_k^H\mathbf{v}_i|^2+\sigma^2}).
\end{equation}
To design $\mathbf{V}$, BS needs to acquire instantaneous downlink CSI, which is conventionally achieved by downlink training and feedback. 
In our scheme, inspired by \cite{Weiyu-MISO}, we propose to bypass explicit channel estimation at the users and directly design the precoder based on downlink received signals since the mutual information between the received pilot and actual channel is larger than that between estimated channel and actual one.
In downlink training stage, BS broadcasts pilot sequence $\mathbf{X} \in \mathbb{C}^{N\times L_{\rm p}}$ of length $L_{\rm P}$ and the received signals $\mathbf{y}_k \in \mathbb{C}^{L_{\rm p}}$ at the $k$-th user can be modeled as:
\begin{equation}
\label{Y}
       \mathbf{y}_k = \mathbf{h}_k^H \mathbf{X} + \mathbf{z}_k,\\
\end{equation}
where $\mathbf{z}_k \in \mathbb{C}^{L_{\rm p}} \sim \mathcal{CN}(\mathbf{0},\sigma^2\mathbf{I})$ is the additive white Gaussian noise and the pilots transmitted at the $l$-th time instant are subject to the transmit power budget $P$ as $\Vert \mathbf{x}_l \Vert_2^2 \leq P$. After the $k$-th user obtain its precoding vector $\mathbf{v}_k$ by its local NN, it subsequently feeds back $\mathbf{v}_k$ to BS in form of $Q$ bits as:  $\mathbf{b}_k = \mathcal{F}(\mathbf{v}_k)$, where function $\mathcal{F}: \mathbb{C}^N \rightarrow \{\pm1\}^{Q}$ represents a certain quantization scheme\footnote{The focus of this work is on how to streamline FDD precoding design and ensure high sum rate performance using heterogeneous multi-vehicle multi-modal sensing. Therefore, for the feedback of precoding vector $\mathbf{v}_k$, we adopt a $B$-bit uniform quantization strategy where a floating-point number is represented by $B$ bits.}.
\begin{figure}[!t]
	\centering
	\includegraphics[width=1\linewidth]{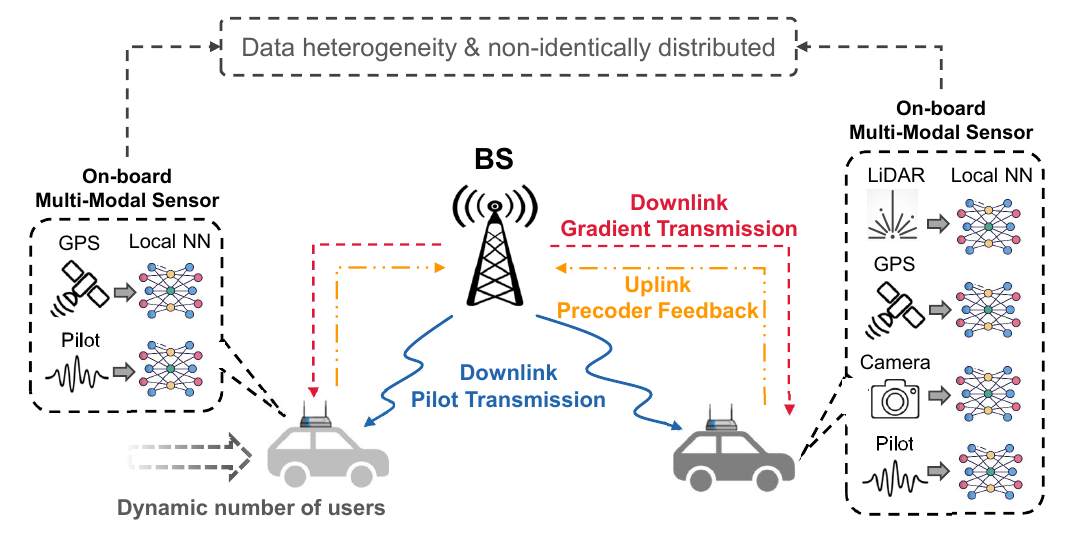}
	\caption{An illustration of the system model.
	\label{wholesystem}}
        \vspace{-1em}
\end{figure}
Given the above established models, the task of maximizing the total data rate within an FDD system can be stated as:
\begin{subequations}
\begin{alignat}{4}
\mathop{\max}\limits_{\mathbf{\Theta}_1,\cdots,\mathbf{\Theta}_K} & \sum_{k=1}^K ~ R_k&{}& 
\label{obj}\\
\mbox{s.t.}\quad
& \mathbf{b}_k = \mathcal{F}(\mathbf{v}_k) = \mathcal{F}(\eta(\mathcal{G}_k(\mathbf{y}_k, \mathbb{M}_k))),  ~~\forall k \\
&\mathbf{V} = [\mathcal{F}^{-1}(\mathbf{b}_1),\cdots,\mathcal{F}^{-1}(\mathbf{b}_K)],\\
&\Tr(\mathbf{V}\mathbf{V}^H) \leq P , 
\end{alignat}
\end{subequations}
where $\mathcal{G}_{k}(\cdot)$ denotes the local NN model of the $k$-th vehicle parameterized by $\mathbf{\Theta}_k, k=1,\cdots,K$. The operator $\eta: \mathbb{R}^{2N}\mapsto\mathbb{C}^N$ defines a mapping that transforms a real-valued vector into a complex-valued vector by taking the first $N$ elements as the real component and the remaining $N$ elements as the imaginary component. Three widely-used sensors are considered in this work: Global Positioning System (GPS), red-green-blue (RGB) cameras, and Light Detection and Ranging (LiDAR). $\mathbb{M}_k$ denotes the set of informative representations obtained from the $k$-th vehicle's available sensory data, i.e. $\mathbb{M}_k = \{ \mathbf{x}_k^{\rm GPS}, \mathbf{x}_k^{\rm RGB},\mathbf{x}_k^{\rm LiDAR}\}$ if the $k$-th vehicle is equipped with three types of sensors. For simplicity,  all subscripts $k$ denote the $k$-th vehicle and will not be restated in the following content.  The optimization objective of this work is the parameter sets of all local models. Fig.~\ref{wholesystem} depicts the schematic of proposed scheme.

\section{Proposed H-MVMM Precoding Scheme Under VFL Framework}
\subsection{Multi-Modal Sensing Preprocessing}
Due to the lack of explicit correlation between raw sensory data and BS precoding, directly feeding raw data into NNs leads to learning difficulties. Hence, we tailored three unique data preprocessing procedures for different sensing modalities to convert them into formats that exhibit explicit correlations with various key characteristics relevant to precoding, which will be introduced in the following subsections. 
\subsubsection{GPS Preprocessing}
\begin{figure}[!t]
	\centering
	\includegraphics[width=1\linewidth]{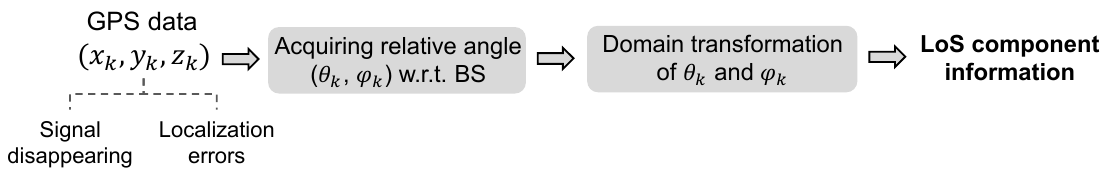}
	\caption{Processing flow of GPS data.
	\label{gps}}
\end{figure}

Intuitively, GPS indicates the rough location of receivers, providing the line-of-sight (LoS) information for BS. However, GPS data is typically presented in coordinate form, which does not align with the data types NNs excel at learning, such as encoded feature vectors. Additionally, BS precoding requires relative angle information between the transceiver rather than the absolute user positions. To address these issues, we devise the following preprocessing scheme tailored for GPS data, as shown in Fig.~\ref{gps}. 

In this work,  without loss of generality, GPS data is assumed to be first processed into Cartesian coordinate form, i.e., $\mathbf{p}_k^t = (x^t_k,y^t_k,z^t_k)$.
Then, we introduce Gaussian noise to raw GPS data to simulate the localization errors in practical applications. Additionally, we account for the issue of GPS signals disappearing due to obstruction by tall buildings. Under such condition, the vehicle continuously generates simulated GPS data based on the most recent available GPS data until the GPS signal is restored:
\begin{equation}
\label{loss}
 \mathbf{p}_k^{(t+\Delta t)} = (x_k^t+v^t_{{\rm x},k}\Delta t,y_k^t+v^t_{{\rm y},k}\Delta t,z_k^t),
\end{equation}
where $v^t_{{\rm x},k}$ and $v^t_{{\rm y},k}$ represent the velocity of vehicle along the x-axis and y-axis at $t$-th time slot which remains constant within $\Delta t$ and are assumed to be accessible by the vehicle.

Based on GPS data and easily accessible BS's location information $\mathbf{p}_{\rm BS} = (x_{\rm BS},y_{\rm BS},z_{\rm BS})$, the vehicle computes the elevation and azimuth angles between itself and BS as:
\begin{gather}
\label{anglecalculation}
       \theta_k = \arctan(\frac{y_k-y_{\rm BS}}{x_k-x_{\rm BS}}),\\
	\phi_k = \arccos(\frac{z_{\rm BS}-z_k}{\Vert \mathbf{p}_k -  \mathbf{p}_{\rm BS} \Vert_2}),
\end{gather}

Since NNs may not effectively learn the differences between multiple raw position data input in coordinate formats, we utilize a high-frequency function~\cite{position encoding} to encode $\theta_k$ and $\phi_k$ into a high-dimensional domain to facilitate NN learning:
\begin{align}
\label{PE}
\gamma(p) = (\sin(2^0\pi p),&\cos(2^0\pi p),
\notag
\\& \cdots,\sin(2^{L-1}\pi p),\cos(2^{L-1}\pi p)),
\end{align}
where $L$ represents the encoding dimension and is set to $5$.

Finally, the LoS component information $\mathbf{x}_k^{\rm GPS} \in \mathbb{R}^{20}$ derived from GPS data is obtained as: $\mathbf{x}_k^{\rm GPS} = (\gamma(\theta_k),\gamma(\phi_k)) $.
\subsubsection{RGB Image Preprocessing}
\begin{figure}[!t]
	\centering
	\includegraphics[width=1\linewidth]{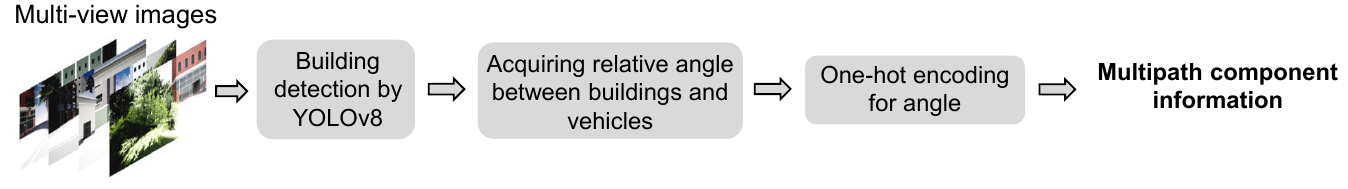}
	\caption{Processing flow of RGB images.
		\label{rgb}}
\end{figure}

RGB images contain rich semantic information, including potential reflectors like buildings contributing to multipath. Yet, without proper preprocessing on the images, NNs may struggle to extract the angular feature of the multipath components that required for BS precoding from raw pixel values. Hence, we devise the following RGB image processing pipeline to convert raw RGB images to coarse multipath angle information as NN input, as shown in Fig.~\ref{rgb}. In this work, to capture a comprehensive view of the vehicle's surroundings, we assume that cameras are equipped on all sides of the vehicle. Let $\mathcal{I}_k = \{\mathbf{I}_{{\rm f},k},\mathbf{I}_{{\rm b},k}, \mathbf{I}_{{\rm l},k}, \mathbf{I}_{{\rm r},k}\}$ be the image set captured by a vehicle, with $\mathbf{I}_{{\rm f},k}$, $\mathbf{I}_{{\rm b},k}$, $\mathbf{I}_{{\rm l},k}$, and $\mathbf{I}_{{\rm r},k}$ being the images captured by cameras mounted on the front, rear, left, and right sides, respectively.

Firstly, we adopt the state-of-the-art object detection model YOLOv8~\cite{yolov8} to detect buildings in the images. Let $f_{\rm{YOLOv8}}(\cdot)$ be the YOLOv8 detector that has been trained on our dataset to accurately detect the buildings. Given an image $ \mathbf{I}_{{\rm f},k} \in \mathbb{R}^{H \times W \times C}$ in image set $\mathcal{I}_k$ as input, where $H$, $W$, and $C$ denote the height, width and the number of color channels, the output bounding boxes and objectness scores of the detected buildings are obtained as:
\begin{gather}
    \label{yolo}
    \{\mathcal{B}^{\rm f}_{1,k},\mathcal{B}^{\rm f}_{2,k},\cdots,\mathcal{B}^{\rm f}_{M,k}\}=f_{\rm{YOLOv8}}(\mathbf{I}_{{\rm f},k}), \\
    \mathcal{B}^{\rm f}_{i,k} = \{u_{i,k},v_{i,k},w_{i,k},h_{i,k},s_{i,k}\}, \enspace i=1,\cdots,M,
\end{gather} 
where $M$ denotes the total number of bounding boxes detected in $\mathbf{I}_{{\rm f},k}$ and $\mathcal{B}^{\rm f}_{i,k}$ denotes the parameter set of the $i$-th bounding box. $u_{i,k}$ and $v_{i,k}$ are the normalized coordinates of the $i$-th bounding box center, respectively; $w_{i,k}$ and $h_{i,k}$ are its normalized weight and height, respectively; and $s_{i,k}$ represents its objectness score. 

Then, we equate the detected buildings to the centers of bounding boxes and calculate their pixel coordinates as: $\mathbf{p}_{i,k}^{\rm I} = (u_{i,k} H, v_{i,k} W, 1)$. Then, we map these pixel coordinates to camera coordinate system to calculate the actual relative angle from the building's center to vehicle by $\mathbf{p}_{i,k}^{\rm C} = \mathbf{K}_{\rm{in}}^{-1} \mathbf{p}_{i,k}^{\rm I}, \enspace i=1,\cdots,M$, where $\mathbf{K}_{\rm{in}}$ represents the intrinsic matrix of camera. Then, $\mathbf{p}_{i,k}^{\rm C}$ is normalized by $\widetilde{\mathbf{p}_{i,k}^{\rm C}} = \frac{\mathbf{p}_{i,k}^{\rm C}}{\mathbf{p}_{i,k}^{\rm C}(3)}$ and the azimuth angle between the building's center and vehicle can be obtained by: $\omega_{i,k} = \arctan(\widetilde{\mathbf{p}_{i,k}^{\rm C}}(1))$.

Following that, we propose to process the azimuth angles of all detected buildings' centers within an image into an informative one-hot vector indicating the angles of all potential multipath relative to vehicle. Note that we only consider detection results with an objectness score larger than $0.5$. Let $\mathbf{r}_{{\rm r},k}$, $\mathbf{r}_{{\rm l},k}$, $\mathbf{r}_{{\rm b},k}$, and $\mathbf{r}_{{\rm f},k}$ be the informative vectors obtained from the right, left, rear, and front cameras, respectively. Taking $\mathbf{r}_{{\rm r},k} \in \mathbb{R}^4$ as an example, it is calculated as follows:
\begin{equation}
    \label{indicator}
    q_{j,i}=\left\{
    \begin{array}{cl}
        0,  &  \lfloor \frac{\omega_{i,k}-\omega_{\text{min}}}{\Delta \omega} \rfloor=j ~ \&\&  ~s_i>0.5, ~i \in [1,M]   \\
        1,  &  \text{otherwise}
    \end{array} \right.,
\end{equation}
\begin{equation}
    \mathbf{r}_{{\rm r},k}(j) = 1- \prod \limits_{i=1}^M q_{j,i},~j=1,2,3,4,
\end{equation}
where $q_{j,i}$ is a logical variable defined for calculating $\mathbf{r}_{{\rm r},k}$, $\omega_{\text{min}}$ is the minimum azimuth angle within the camera's horizontal field of view, and $\Delta \omega$ is the angle interval. Subsequently, the indicator vector $\mathbf{r}_k$ of all potential multipath angles relative to vehicle in the surrounding environment is derived by $\mathbf{r}_k = (\mathbf{r}_{{\rm r},k},\mathbf{r}_{{\rm l},k},\mathbf{r}_{{\rm b},k},\mathbf{r}_{{\rm f},k}) \in \mathbb{R}^{16}$. Note that the orientation $\beta_{M,k}$ of vehicle is another crucial element that allows BS to learn the absolute angle information about the multipath, which can be easily measured by vehicular sensors like inertial measurement units. Similar to the processing of GPS data, we also encode $\beta_{M,k}$ using the high-frequency function with the encoding dimension set to $10$.  

Finally, the multipath component information offered by RGB images is formed as: $\mathbf{x}_k^{\rm RGB} = ( \mathbf{r}_k,\gamma(\beta_{M,k}) )  \in \mathbb{R}^{36}$.

\subsubsection{LiDAR Preprocessing}
\begin{figure}[!t]
	\centering
	\includegraphics[width=1\linewidth]{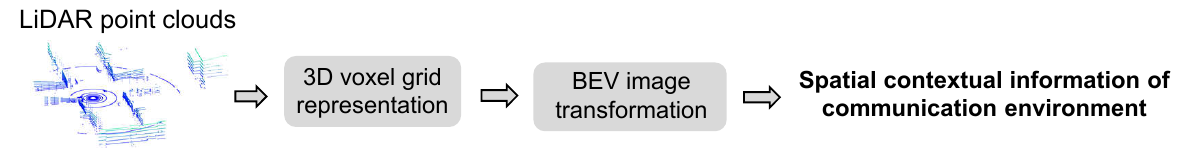}
	\caption{Processing flow of LiDAR point clouds.
		\label{lidar}}
\end{figure}

LiDAR point clouds collected by multiple vehicles collaboratively depict the spatial structural information of the communication environment, potentially assisting BS in modeling the complex relationships among users through NNs. However, inputting raw LiDAR data into NNs results in extremely high complexity, leading to prolonged NN inference time. To mitigate this, we perform the following lightweight processing, as depicted in Fig.~\ref{lidar}.

Firstly, we represent the entire point cloud through a 3D voxel grid $\mathbf{Z}_k \in \mathbb{R}^{L_{\rm x}\times L_{\rm y}\times L_{\rm z}}$. If a voxel in $\mathbf{Z}_k$ contains at least one point of the entire point cloud, it is occupied and marked as $1$. Otherwise, the voxel's value is set to $0$. Next, a bird's eye view  (BEV) image $\mathbf{X}^{\rm LiDAR}_k \in \mathbb{R}^{L_{\rm x}\times L_{\rm y}}$ is formed based on $\mathbf{Z}_k$ by overlaying the values of all voxels in the z-direction and treating the result as the pixel value. Finally, $\mathbf{X}^{\rm LiDAR}_k$ contains sufficient structural information of the vehicle's surronding environment and is regarded as the input of the LiDAR branch.

Note that the above preprocessing is simpler compared to that designed for GPS data and RGB images. This is because LiDAR raw data has already shown a significant correlation with the precoding task, enabling NNs to extract effective features from it without the need for complex processing. On the other hand, raw GPS data and RGB images have no explicit relationship with the wireless channel. Only through intricate preprocessing can they be transformed into informative representations directly correlated to precoding task. 

\subsection{Local Training at Vehicles}
A vehicle's local model $\mathcal{G}_{k}(\cdot)$ is composed of several uni-modal feature extraction NNs and a multi-modal feature integration NN. Let $\mathcal{G}^{\rm G}(\cdot)$, $\mathcal{G}^{\rm R}(\cdot)$, $\mathcal{G}^{\rm L}(\cdot)$, and $\mathcal{G}^{\rm P}(\cdot)$ be the feature extraction NN designed for GPS, RGB images, LiDAR, and pilots. Then, the multi-modal features are extracted as:
\begin{subequations}
    \begin{align}
    &\mathbf{m}_k^{\rm G} = \mathcal{G}^{\rm G}(\mathbf{x}^{\rm GPS}_k), &\mathcal{G}^{\rm G}: \mathbb{R}^{20} \mapsto \ \mathbb{R}^{L_{\rm G}},\label{g}\\
    &\mathbf{m}_k^{\rm R} = \mathcal{G}^{\rm R}(\mathbf{x}^{\rm RGB}_k), &\mathcal{G}^{\rm R}: \mathbb{R}^{36} \mapsto \ \mathbb{R}^{L_{\rm R}}, \label{r}\\
    &\mathbf{m}_k^{\rm L} = \mathcal{G}^{\rm L}(\mathbf{X}^{\rm LiDAR}_k), &\mathcal{G}^{\rm L}: \mathbb{R}^{L_{\rm x} \times L_{\rm y}} \mapsto \ \mathbb{R}^{L_{\rm L}},\label{l}\\
    &\mathbf{m}_k^{\rm P} =  \mathcal{G}^{\rm P}(\mathbf{y}_k), &\mathcal{G}^{\rm P}: \mathbb{R}^{2L_{\rm P}} \mapsto \ \mathbb{R}^{L_{\rm S}}, \label{p} 
    \end{align}
\end{subequations}
where $\mathbf{m}_k^{\rm G}$, $\mathbf{m}_k^{\rm R}$, $\mathbf{m}_k^{\rm L}$, and $\mathbf{m}_k^{\rm P}$ denote the uni-modal features obtained from GPS, RGB image, LiDAR, and received pilots, respectively. Assume that the $k$-th vehicle is equipped with three sensors. Then, the vehicle concatenates the uni-modal features as $\mathbf{m}_k^{\rm I} = \mathbf{m}_k^{\rm G} \oplus \mathbf{m}_k^{\rm R} \oplus \mathbf{m}_k^{\rm L} \oplus \mathbf{m}_k^{\rm P}$, where $\oplus$ denotes the concatenation operation. Finally, a multi-modal feature integration NN $\mathcal{G}^{\rm I}(\cdot)$ is used to output the precoding vector for the $k$-th user leverging $\mathbf{m}_k^{\rm I}$:
\begin{equation}
\label{v}
    \mathbf{v}_k = \eta(\mathcal{G}^{\rm I}(\mathbf{m}_k^{\rm I})), \enspace \enspace \mathcal{G}^{\rm I}: \mathbb{R}^{L_{\rm G}+L_{\rm R}+L_{\rm L}+L_{\rm S}} \mapsto  \mathbb{R}^{2N_{\rm t}}.
\end{equation}

\subsection{Training Procedure of the Proposed VFL-Based Precoding}
The training steps of the proposed scheme is illustrated in Algorithm 1. Specifically, in the $j$-th iteration, each vehicle first calculates its local model output $\mathbf{v}_k = \mathcal{G}_k(\mathbf{y}_k, \mathbb{M}_k))$  and feeds back it to BS. During the training stage, we assume that more quantization bits can be used for uplink feedback to ensure network performance. Utilizing all $\mathbf{v}_k$, BS evaluates the training loss as: 
\begin{equation}
\label{loss}
\mathcal{L} = -\sum_{k=1}^K (R_k + \lambda^{(R_{\rm T}-R_k)}).
\end{equation}
In Eq.~\eqref{loss}, except for the reciprocal of the sum rate commonly used in existing works, there is an additional specially designed loss term. 
Its role is to reflect the gap between each user's achievable rate $R_k$ and the threshold $R_{\rm T}$. 
During the initial stage of network training, a large gap $(R_{\rm T}-R_k)$ results in a large derivative, effectively accelerating the convergence speed of smaller local models.

Then, BS calculates the gradients $\frac{ \partial \mathcal{L} }{ \partial \mathbf{v}_k }$ for each vehicle and transmits them back to vehicles. Subsequently, each vehicle computes the gradient of its local model $\mathbf{\Theta}_k^j$ and updates it as shown in the eighth and ninth lines of Algorithm 1, where $\eta_j$ represents the learning rate in the $j$-th iteration. The above process iterates until the entire NN converges. 

\begin{figure}[!t]
    \label{alg}
    \renewcommand{\algorithmicrequire}{\underline{\textbf{Input:}}}
    \renewcommand{\algorithmicensure}{\underline{\textbf{Output:}}}
    \begin{algorithm}[H]
        \caption{Training Steps of VFL-based Precoding}
        \begin{algorithmic}[1]
            \REQUIRE Available multi-modal data of vehicles $\{\mathbf{y}_k, \mathbb{M}_k\}_{k=1}^K$.   
            \ENSURE Precoding matrix $\mathbf{V}$.  
            \STATE Vehicles $1,2,\cdots,K$ initialize $\mathbf{\Theta}_k=\mathbf{\Theta}_k^{0}$. 
            \FOR{\text{each iteration} $j=1,2,\cdots $ } 
             \FOR{\text{each vehicle} $k=1,2,\cdots,K$ \text{in parallel}}
            \STATE Compute $\mathbf{v}_k = \eta(\mathcal{G}_k(\mathbf{y}_k, \mathbb{M}_k))$ and feed it back to BS;
            \ENDFOR
            \STATE BS collects vehicle's feedback bits and reconstructs $\mathbf{V}$. 
            \STATE BS computes and sends $[\frac{ \partial \mathcal{\mathcal{L}} }{  \partial\mathbf{v}_k  }]_{k=1}^K$ to each vehicle.  
            \FOR{\text{each vehicle} $k=1,2,\cdots,K$ \text{in parallel}}
            \STATE  Vehicle $k$ computes $\nabla_{\mathbf{\Theta}_k^{j}}\mathcal{L}= \frac{ \partial \mathcal{L} }{ \partial \mathbf{v}_k } \frac{ \partial\mathbf{v}_k }{ \partial \mathbf{\Theta}_k^{j}}$; 
             \STATE  Vehicle $k$ updates $\mathbf{\Theta}_k^{j} = \mathbf{\Theta}_k^{j-1}-\eta_j \nabla_{\mathbf{\Theta}_k^{j}}\mathcal{L} $; 
            \ENDFOR
            \ENDFOR
        
        \end{algorithmic}
    \end{algorithm}
\end{figure}
%
\subsection{Online Training Strategy for Variable User Number}
Most existing DL approaches pre-train and store multiple NNs with varying input/output sizes to accommodate changes in user number. However, this strategy proves impractical in our scenario, where both the user count and their sensor configurations are variable. For instance, when a new user joins, existing users would need to pre-download numerous models to adapt to the new user's potential local model configurations. Specifically, with $K=6$, accommodating $N_{\rm sensor}$ potential sensor configurations for the new user would require storing up to $N_{\rm sensor} \times K = (C_{3}^{1}+C_{3}^{2}+C_{3}^{3})\times 6 = 42$ pre-trained models in advance.

To address this inflexibility, we propose an online training strategy within the VFL framework. During the online inference stage, when the number of vehicles changes, new vehicles first download a randomly initialized model from BS and subsequently train the corresponding NN branches from scratch based on their available sensors. Concurrently, existing vehicles will retrain their local pre-trained models.  The training procedure adheres to Algorithm 1, with the BS utilizing downlink channel estimation for loss calculation. This online fine-tuning strategy enables the overall NN to converge in a short number of epochs, allowing the system to quickly re-enter the inference stage.
\section{Numerical Results}
\subsection{Experiment Setting}
\subsubsection{Dataset}
To showcase the environment characterization abilities of multi-vehicle multi-modal sensing in complex communication environments, we adopt the city block scenario from the M$^3$SC dataset~\cite{M3C}. This scenario features packed buildings and many non-line-of-sight communication conditions. The scenario size is  $190$m  $\times$ $135$m. The BS is equipped with $N_{\rm{v}} \times N_{\rm h} = 16 \times 8$ antennas, with the height set to $z_{\rm BS} = 9$m. The downlink and uplink carrier frequency is $4.95$ GHz and $4.85$ GHz, respectively. The mean and standard deviation of Gaussian noise added to the x and y coordinates of GPS data is set to $0$ and $5$, respectively. 

\subsubsection{Implementation Details}
\begin{figure}[!t]
	\centering
	\includegraphics[width=1\linewidth]{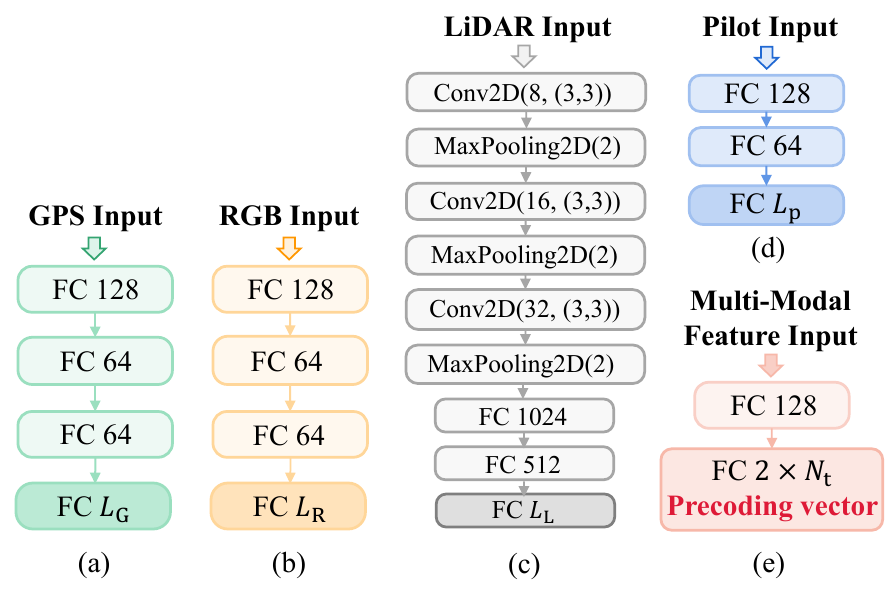}
	\caption{Illustration of NN architectures for: (a) GPS branch; (b) RGB image branch; (c) LiDAR branch; (d) received pilots branch; (e) multi-modal feature integration network.
	\label{NN}}
\end{figure}

We use the adaptive moment estimation (ADAM) as the optimizer and initialize the learning rate to $1 \times 10^{-4}$ for all schemes. We train different schemes with a batch size of $32$ until the validation loss stabilizes. The architectures of NNs customized for different sensing modality and the multi-modal feature integration network are depicted in Fig.~\ref{NN}. The sizes of multi-modal features are set to $L_{\rm S}=128$, $L_{\rm G}=256$, $L_{\rm R}=256$, and $L_{\rm L}=512$, respectively. The threshold $R_{\rm T}$ is set to $0.3$ and the base $\lambda$ is set to $10$ in the loss function, which has been experimentally proven to be suitable for optimizing the training of local models. Unless specifically mentioned, the length of pilot sequence and number of quantization bits utilized by all VFL-based schemes is set to $L_{\rm P}=8$ and $B=2$, respectively. In the H-MVMM scheme, all vehicles have the pilot modality, with $5$ out of $7$ vehicles equipped with GPS, $3$ vehicles equipped with RGB cameras, and $3$ vehicles equipped with LiDAR.

\addtolength{\topmargin}{0.03in}

\begin{table}[!t]
	\setlength{\abovecaptionskip}{0.1cm} 
	\renewcommand\arraystretch{0.3} 
	\centering
	\caption{Sum rate performances of different schemes with respect to $K$ for SNR=$30$dB.}
        \resizebox{1\linewidth}{!}{
	\label{sumrate1}
	\begin{tabular}{@{\hspace{1.2em}}c@{\hspace{1.2em}}|@{\hspace{1.2em}}c@{\hspace{1.2em}}}
		\toprule[0.35mm]
		{\textbf{Scheme}}  &\makecell[c]{\textbf{Sum rate performance} (bps/Hz) \\ (K=$3/4/5/6/7$)}  \\
		\midrule[0.15mm]
  		\makecell[c]{VFL-Uni-Pilot scheme ($L_{\rm P}=8$)\\ VFL-Uni-Pilot scheme ($L_{\rm P}=32$)} & \makecell[c]{$27.3$~/~$36.3$~/~$45.2$~/~$54.4$~/~$65.9$ \\ $28.4$~/~$37.4$~/~$47.0$~/~$56.5$~/~$69.3$} \\ 
		\midrule[0.15mm]
		\makecell[c]{VFL-Pilot-LiDAR scheme}  & {$33.1$}~/~{$42.2$}~/~{$54.1$}~/~{$66.9$}~/~{$80.2$} \\ 
		\midrule[0.15mm]			
		\makecell[c]{VFL-Pilot-RGB scheme} 	& $31.1$~/~$39.6$~/~$50.4$~/~{$60.9$}~/~$72.4$  \\
		\midrule[0.15mm]			
		  \makecell[c]{VFL-Pilot-GPS scheme}  & $29.1$~/~$37.3$~/~$45.5$~/~{$56.9$}~/~$68.1$   \\
		\midrule[0.15mm]		
		\makecell[c]{VFL-H-MVMM scheme} & $30.5$~/~$40.0$~/~$51.2$~/~{$63.1$}~/~$75.2$  \\ 
             \midrule[0.15mm]		
		\makecell[c]{{\color{black}\textbf{WMMSE w/ GT CSIT scheme}}} & \textbf{{35.2}}~/~\textbf{{45.0}}~/~\textbf{{56.6}}~/~\textbf{{68.7}}~/~\textbf{81.6 } \\ 
		\bottomrule[0.35mm]		
	\end{tabular}
 }
\vspace{-1em}
\end{table}

\subsubsection{Baseline Methods}
We consider using WMMSE~\cite{WMMSE} and ZF method~\cite{ZF} with ground truth CSI as baseline methods (represented by w/ GT CSIT), as WMMSE can achieve a locally optimum performance.
In addition, a BS-NN scheme utilizing CSI estimates (CE) as input is designed, with the normalized mean squared error being $-30$dB. Since the purpose of pilot transmission is to obtain accurate CSI, the BS-NN can be considered as the upper bound of Uni-Pilot scheme.
\subsection{Performance Comparisons to Baseline Methods}
In the following experiments, we set the signal-to-noise-ratio (SNR) as  $\text{SNR}=10\log_{10}(\frac{P}{\sigma^2})=30$ dB unless otherwise specified.
To showcase the role of a certain sensing modality, we assume that all vehicles are equipped with the same type of sensor and evaluate the sum rate performances in Table.~\ref{sumrate1}. The Pilot-LiDAR scheme achieves a comparable performance with WMMSE method w/ GT CSIT thanks to the collaborate characterization of the communication environment by all on-board LiDARs, demonstrating its greatest role across all sensing modalities. Despite the presence of localization errors and signal disappearing issues, the LoS component information provided by GPS still brings an evident performance enhancement. The role of RGB images falls between LiDAR and GPS, while the performance of the H-MVMM scheme depends on the vehicles' available sensors.

\begin{figure}[!t]
	\centering
	\includegraphics[width=1\linewidth]{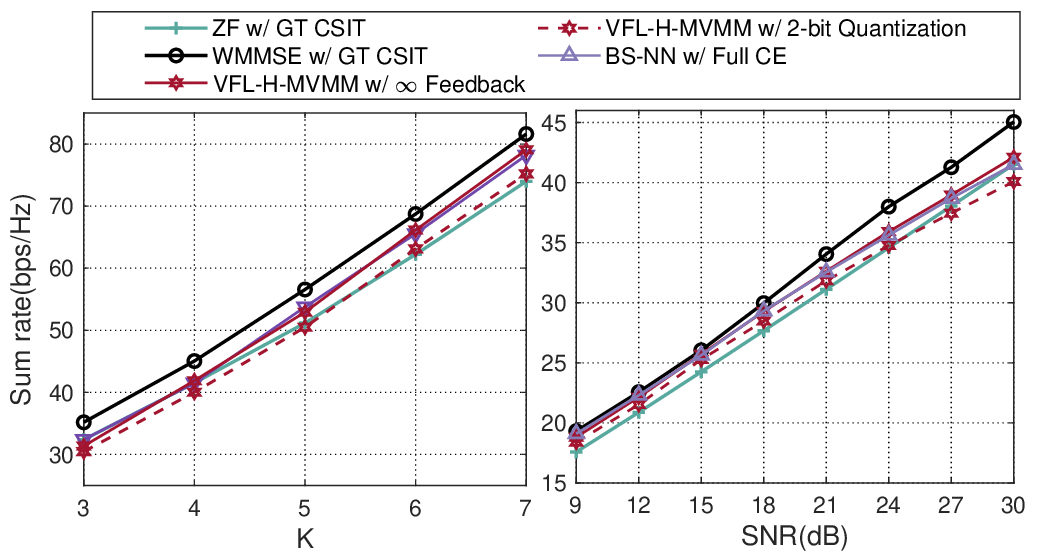}
	\caption{Performances of baseline methods and H-MVMM scheme: (a) vs. $K$ at SNR=$30$dB; (b) vs. SNR at $K=4$.
		\label{rate}}
\end{figure}

Fig.~\ref{rate}(a) compares the sum rate achieved by baseline methods and the H-MVMM scheme. The H-MVMM scheme w/ limited feedback can approach the performance of ZF w/ GT CSIT when $K < 5$ and surpass it as $K$ increases. Furthermore, as $K$ increases, the H-MVMM scheme w/ sufficiently large $B$ approximates the performance of WMMSE w/ GT CSIT and outperforms BS-NN w/ full CE.
Fig.~\ref{rate}(b) depicts the performances of different schemes against different SNRs. We can see that the advantage of the H-MVMM scheme w/ limited feedback is more evident in low SNR regimes. Its performance closely approaches those of WMMSE and BS-NN. Despite the performance gap between H-MVMM and WMMSE, the H-MVMM scheme enjoys significantly lower complexity. The time complexity of WMMSE in one iteration is $\mathcal{O}(K^2{N^3})$ \cite{WMMSE} while that of H-MVMM is on the order of $\mathcal{O}(L_{\rm P}N)$. The H-MVMM scheme achieves a constant time complexity increase with respect to $N$ and $L_{\rm P}$ ($L_{\rm P} \ll N$).
%

\subsection{Towards Adaptability to Change in User Number}
\begin{figure}[!t]
	\centering
	\includegraphics[width=1\linewidth]{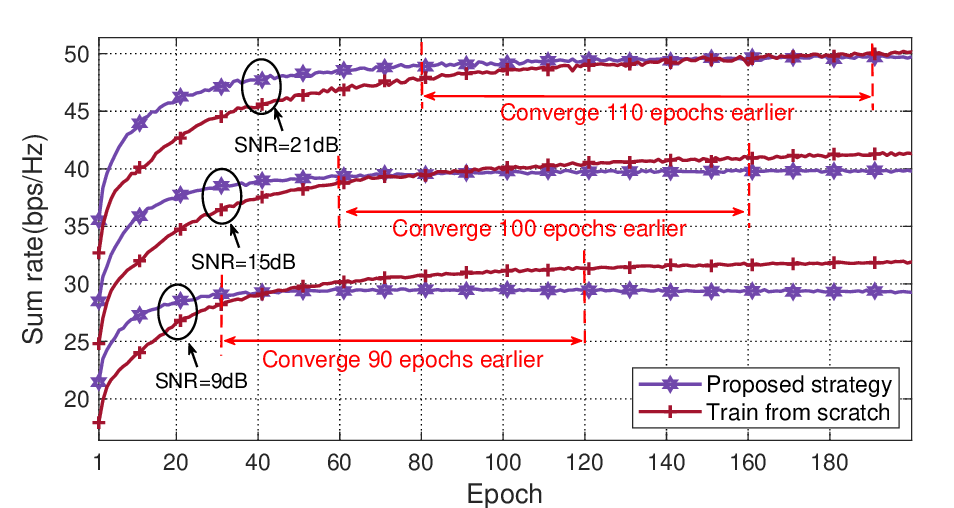}
	\caption{Sum rate achieved by different strategies versus number of training epochs when $K=6$ and SNR$=[9, 15, 21]$dB.
	\label{online}}
\end{figure}

Fig.~\ref{online} depicts the sum rate achieved by the proposed online training strategy when $K$ changes from $5$ to $6$ under different SNRs, alongside that of offline training a NN for $K=6$ from scratch. It can be seen that the proposed strategy achieves a faster convergence speed, saving at least $90$ epochs. Though the time required for re-convergence is inevitably longer than directly switching to a pre-trained model, the proposed strategy holds practical applicability thanks to its capability to dynamically adapt to any user number and sensor configurations. Regarding sum rate performance, since BS can only adopt CSI estimates rather than ground truth to compute loss, models trained by the proposed strategy inevitably faces performance degradation.
\subsection{Comparisons to Centralized Learning}
We present in Table~\ref{setup} the differences in implementing the proposed scheme within VFL and CL frameworks. In terms of data consumption required by network training, VFL-based schemes only require users to transmit a precoding vector of size $N_{\rm t} \times 1$ to BS in each epoch via uplink, thus requiring only a very small amount of data transmission. Specifically, the amount of data transmission in VFL is given by $W_{\rm FL} = N_{\rm epoch} \times K \times 0.5$KB, where $N_{\rm epoch}$ is the number of training epochs. Conversely, centralized learning necessitates gathering training data collected by all users and leads to enormous data transmission costs, which carries risks of infringing on user privacy and is impractical for real-world applications. 

Despite the need to collect vast amounts of training data, the advantage of centralized learning is that BS can directly obtain the precoding matrix and avert the quantization errors caused by limited feedback. Nonetheless, as shown in Table~\ref{setup}, the performance losses resulting from quantization errors are minimal. Therefore, it can be concluded that VFL is the more suitable learning framework in the considered scenario.

\begin{table}[!t]
	\setlength{\abovecaptionskip}{0.1cm} 
	\renewcommand\arraystretch{0.05} 
	\centering
	\caption{Comparison of different schemes under VFL and CL frameworks when $K$=7.}
 \resizebox{1\linewidth}{!}{
	\label{setup}
         \begin{tabular}{@{\hspace{0.6em}}c@{\hspace{0.6em}}|@{\hspace{1em}}c@{\hspace{1em}}|@{\hspace{1em}}c@{\hspace{1em}}}
		\toprule[0.35mm]
		{\textbf{Scheme} (VFL/CL)}  & {\textbf{Dataset volume} (MB)} &{\textbf{Sum rate}  (bps/Hz)}  \\
		\midrule[0.15mm]
		\makecell[c]{Pilot-LiDAR scheme} & \underline{$0.7$}/$532.0$ & \underline{$80.2$}/$83.7$ \\ 
		\midrule[0.15mm]			
		\makecell[c]{Pilot-RGB scheme} & \underline{$2.1$}/$98.0$	& \underline{$72.4$}/$75.9$  \\
		\midrule[0.15mm]			
		  \makecell[c]{Pilot-GPS scheme}  & \underline{$2.8$}/$85.1$ & \underline{$68.1$}/$72.3$   \\
		\midrule[0.15mm]		
		\makecell[c]{H-MVMM scheme}& \underline{$2.8$}/$282.8$ & \underline{$75.2$}/$79.1$  \\ 		
		\bottomrule[0.35mm]		
	\end{tabular}	
 }
\end{table}



\section{Conclusions}
This paper presents a VFL-based precoding approach for FDD systems that reduces both pilot overhead and computational complexity. By employing carefully designed data preprocessing methods and a tailored VFL training procedure, our heterogeneous multi-vehicle, multi-modal sensing effectively extracts a variety of precoding-related features, enhancing the function of pilot transmission. Additionally, we introduce an online training strategy that allows the proposed scheme to dynamically adapt to fluctuations in user numbers with minimal adaptation times. Simulation results demonstrate that our approach, utilizing significantly shorter pilot sequences, achieves performance closely comparable to classical optimization schemes under perfect CSI.


\end{document}